\documentclass[twocolumn]{aastex631}
\usepackage{bm}
\usepackage{amsmath}
\graphicspath{{./}{figures/}}

\begin{document}

\title{Polar orbits around the newly formed Earth-Moon binary system}

\author[0000-0003-2270-1310]{Stephen Lepp}

\author[0000-0003-2401-7168]{Rebecca G. Martin}

\author[0000-0003-0412-760X]{Stanley A. Baronett}

\affiliation{Nevada Center for Astrophysics, University of Nevada, Las Vegas, 4505 S. Maryland Pkwy., Las Vegas, NV 89154, USA}
\affiliation{Department of Physics and Astronomy,University of Nevada, Las Vegas, 4505 S. Maryland Pkwy., Las Vegas, NV 89154, USA}

\begin{abstract}
We examine the dynamics and stability of circumbinary particles orbiting around the Earth-Moon binary system. The moon formed close to the Earth (semi-major axis $a_{\rm EM}\approx 3\,\rm R_\oplus$) and expanded through tides to its current day semi-major axis ($a_{\rm EM}= 60\,\rm R_\oplus$).   Circumbinary orbits that are polar or highly inclined to the Earth-Moon orbit are subject to two competing effects: (i) nodal precession about the Earth-Moon eccentricity vector and (ii)  Kozai-Lidov oscillations of eccentricity and inclination driven by the Sun. 
While we find that there are no stable polar orbits around the  Earth-Moon orbit with the current day semi-major axis, polar orbits were stable immediately after the formation of the Moon, at the time when there was a lot of debris around the system, up to when the semi-major axis reached about $a_{\rm EM}\approx 10\,\rm R_\oplus$. We discuss implications of polar orbits on the evolution of the Earth-Moon system and the possibility of polar orbiting moons around exoplanet-moon binaries.
\end{abstract}


\section{Introduction}
\label{sec:intro}

Our Moon has a relatively large mass of about 1.2\% of  the mass of the Earth and currently orbits at a semi-major axis of $a_{\rm EM}=60\,\rm R_\oplus$ with an orbital inclination of about $5^\circ$ to the ecliptic plane.  The Moon is gradually receding from the Earth as a result of tidal friction caused by lunar tides. Therefore, when the Moon formed, it was much closer to the Earth, and the Earth was rotating faster \citep{Darwin1879}. 

Our Moon is thought to have formed from a late giant impact of a Mars-sized object with the proto-Earth \citep{Hartmann1975,Cameron1976}. During the impact, a circumterrestrial debris disc formed from which the Moon accumulated \citep{Benz1986, Cameron2000, Canup2001,Wada2006}.
The minimum distance that the Moon can possibly form around the Earth is the Earth's Roche limit at $a_{\rm EM}=2.9\,\rm R_\oplus$ for lunar density material. The Moon is thought to have formed just outside of this radius at $a_{\rm EM}=3-5\,\rm R_\oplus$ \citep{Salmon2012, Cuk2012}.

At the time of Moon formation, the Earth was molten and tidal dissipation was weak. However,  as the magma ocean began to solidify, tidal dissipation increased. At the end of the solidification of the magma ocean at a time of a few $10^4\,\rm yr$ since formation, the semi-major axis was around $a_{\rm EM}=7-9\,\rm R_\oplus$ \citep{Korenaga2023}. The water oceans formed later and, at this stage, tidal dissipation increased again.
Currently, ocean tides are more important than solid Earth tides and these ocean tides are sensitive  to the rotation rate along with ocean properties such as the depth and volume \citep{Webb1982, Motoyama2020}.

Since its formation, the semi-major axis of the Moon has expanded more significantly than a moon around any other planet in the solar system as a result of tidal interactions \citep{Canup2004b}. Initially the Moon was in an orbit with a semi-major axis that was at least 15 times smaller and the orbit was inclined to the equatorial plane by at least $10^\circ$ when the semi-major axis was around $a_{\rm  EM}=10\,\rm R_\oplus$ \citep{Goldreich1966, Touma1994}. The origin of this initial inclination is not well explained since the forming moon would be expected to form aligned to the equatorial plane of the Earth \citep[e.g.][]{Cuk2016}. However, it could be explained through either the impact itself, disc torques or solar resonances \citep{Ward2000,Canup2004b}.

The Moon forming impact created large amounts of debris, some of which was accreted on to the Earth or the Moon, while some was ejected from the Earth-Moon system \citep[e.g.][]{Jackson2012}. In this work, we study the orbital dynamics of circumbinary Earth-Moon material.  Since the orbital eccentricity of the Earth-Moon orbit is non-zero, there are two types of nodal precession of misaligned orbits around the system. {\it Circulating} orbits are those in which the orbit of a particle precesses about the angular momentum vector of the Earth-Moon binary. {\it Librating} orbits occur around eccentric binaries when the initial inclination of the particle is large. The particle precesses about the binary eccentricity vector \citep{Verrier2009, Farago2010,Doolin2011}.  In the absence of the moon, such highly inclined obits would be unstable to Kozai-Lidov \citep[KL,][]{Kozai1962,Lidov1962} oscillations of inclination and eccentricity. However, the libration driven by an inner binary suppresses the KL oscillations and allows particles to exist on highly inclined orbits \citep{Verrier2009,Martin2022}.

In this work we examine how the dynamics and stability of circumbinary material around the Earth-Moon binary has changed since the formation of the Moon. In Section~\ref{setup} we describe the set up for our simulations. In Section~\ref{phase} we examine the dynamics of circumbinary material around the Earth-Moon binary. In Section~\ref{stability} we consider where stable circumbinary material is possible around the Earth-Moon binary. In Section~\ref{tides} we consider the effect of tides and we draw our conclusions in Section~\ref{conc}.

\section{Simulation set up}
\label{setup}

The simulations in this paper make use of the {\sc rebound} $n$-body code \citep{rebound}. The simulations are integrated using WHFast, a symplectic Wisdom-Holman integrator \citep{reboundwhfast,wh}. 

Our standard parameters model the  Earth-Moon system orbiting the Sun at semi-major axis $a_{\rm ES}=1 \,\rm au$ in a circular orbit ($e_{\rm ES}=0$). The Earth-Moon orbit has eccentricity of $e_{\rm EM}=0.055$ and is coplanar to the orbit around the Sun.   The masses of the Sun, Earth and Moon are all set to their current value. We vary the semi-major axis of the Earth-Moon binary in the range of $a_{\rm EM}=3-60\,\rm R_\oplus$, where $\rm R_{\oplus}$ is the current Earth radius. This range covers the approximate semi-major axis  of the Earth-Moon system at its formation to the current value. We have simplified the system for easier analytical analysis, but we have run tests with the current inclination of the EM orbit relative to the SE orbital plane ($5^\circ$) and the eccentricity of the orbit about the Sun ($e_{\rm ES}=0.017$) and find very little difference in our results. 

We examine test particle orbits  around the Earth-Moon system.  These orbits are initially circular with semi-major axis $R$ and with an inclination to the plane of the Earth-Moon system given by 
\begin{equation}
i = \cos^{-1}(\hat{\bm{l}}_{\rm EM}\cdot \hat{\bm{l}}_{\rm t})\,,
\end{equation}
where $\hat{\bm{l}}_{\rm EM}$ and $\hat{\bm{l}}_{\rm t}$ are the unit vectors in the direction of the angular momentum of the Earth-Moon system and the test particle respectively.
The nodal  phase angle of the test particle is 
defined as the angle measured relative to 
the eccentricity vector of the Earth-Moon system and is 
given by 
\begin{equation}
        \phi = \tan^{-1}\left(\frac{\hat{\bm{l}}_{\rm t}\cdot (\hat{\bm{l}}_{\rm EM}\times 
    \hat{\bm{e}}_{\rm EM})}{\hat{\bm{l}}_{\rm t}\cdot \hat{\bm{e}}_{\rm EM}}\right) + 90^\circ 
\end{equation}
\citep{Chen2019,Chen2020e}, where  $\hat{\bm{e}}_{\rm EM}$ is the  unit eccentricity vector of the Earth-Moon binary. We start the test particle orbits with the initial angular momentum in the plane defined by the Earth-Moon angular momentum  vector and its eccentricity vector.  Thus we have an initial inclination $i$ and the initial phase angle is $\phi=90^\circ$.

The simulations are run initially without tidal effects but we  explore the effect of tides in Section~\ref{tides}. 
The effect of tides on circumbinary orbits is most important at small semi-major axis of the Earth-Moon system.  There the tides cause significant apsidal precession of the Earth-Moon orbit and affect the librating orbits of test particles around the system.

\section{Orbital dynamics of circumbinary Earth-Moon material}
\label{phase}

\begin{figure*}
\centering
\hspace{-0.0in}\includegraphics[width=1.3\columnwidth]{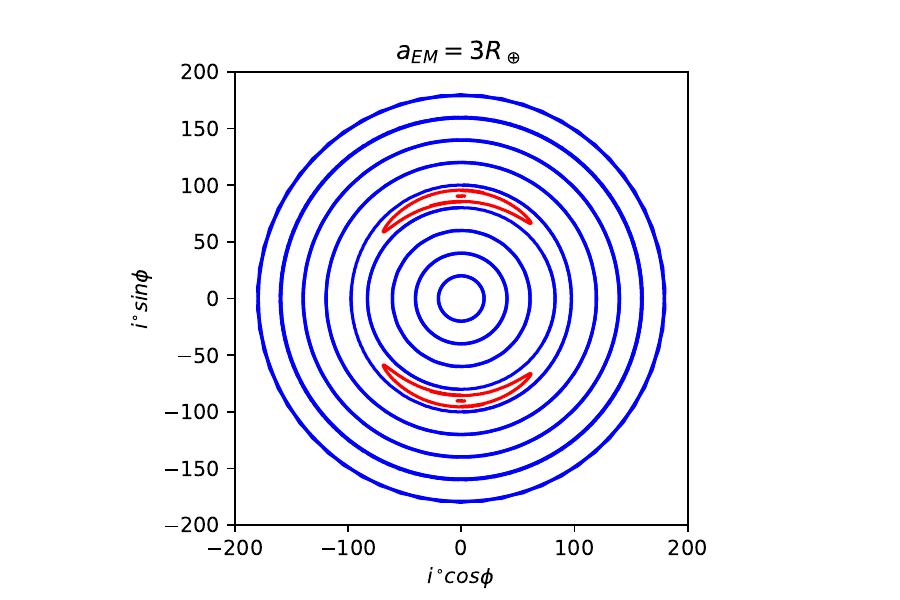}\hspace{-1.8in}\includegraphics[width=1.3\columnwidth]{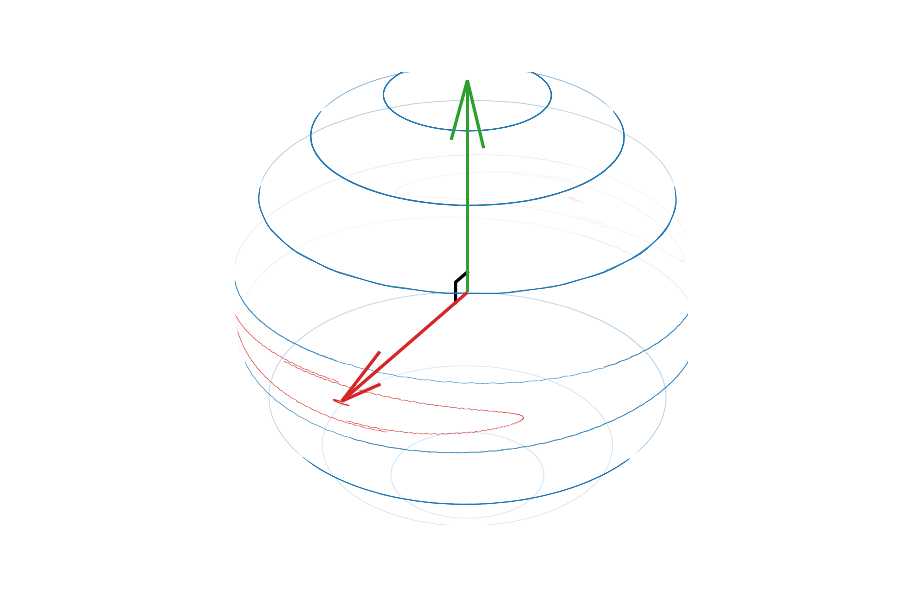}
\includegraphics[width=1.3\columnwidth]{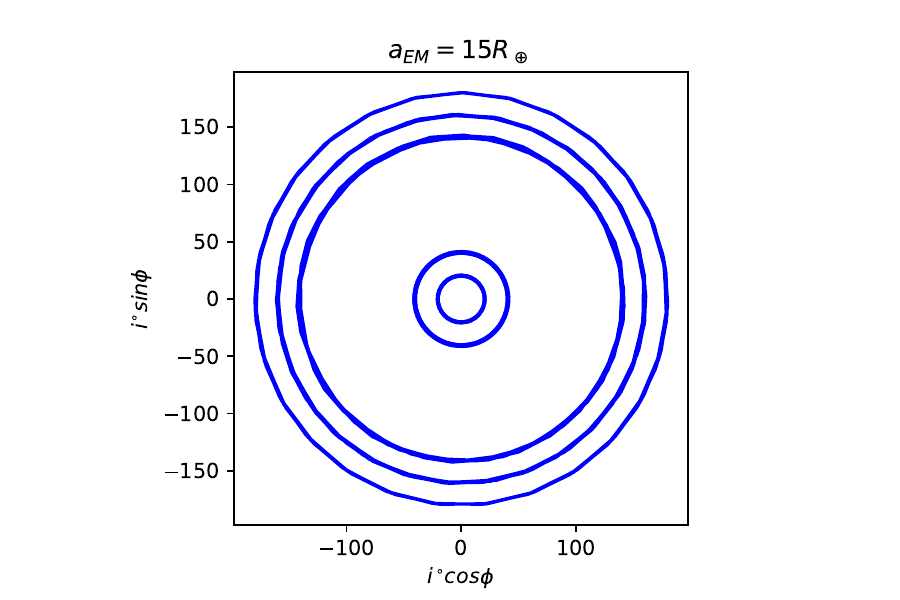}\hspace{-1.8in}
\includegraphics[width=1.3\columnwidth]{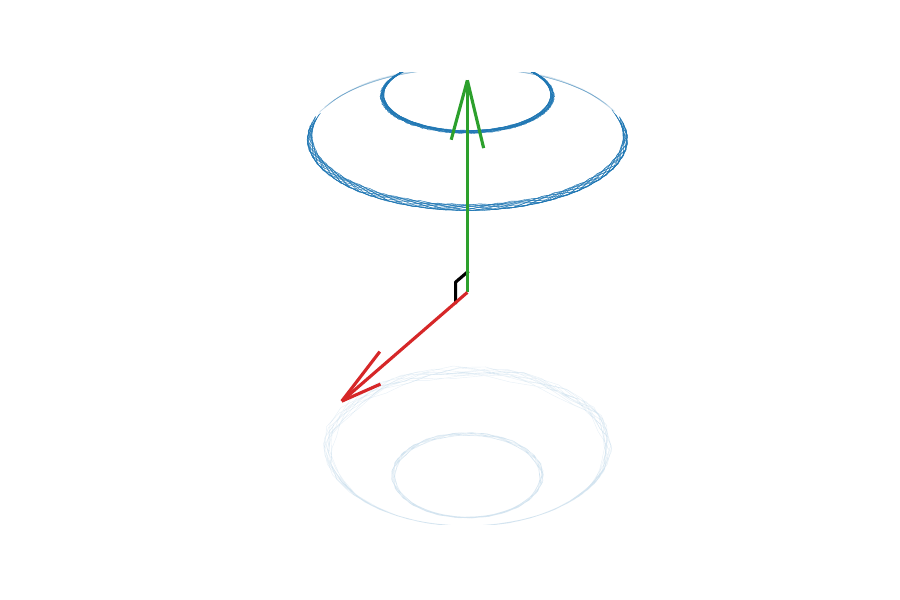}
    \caption{
    Top left: The $i\cos \phi-i\sin \phi$ phase diagram for test particles orbiting around the Earth-Moon binary with semi-major axis  $a_{\rm EM}=3\,\rm R_\oplus$. The particles have a semi-major axis of $R=4 a_{\rm EM}$.  The blue circulating orbits are at initial inclinations of 20, 40, 60, 80, 100, 120, 140, 160 and 180 and the red librating orbits are at initial inclinations of 85, 90 and 95.  
Top right: The same as the top left but the path of the angular momentum vector of the particle is plotted on a unit sphere.  The green  arrow is the unit angular momentum vector and the red arrow is the unit eccentricity vector  for the Earth-Moon system.
    Bottom panels: Same as the top panels but for  $a_{\rm EM}=15\,\rm R_\oplus$ with a particle semi-major axis of $R= 4 \,a_{\rm EM}$.   }
    \label{fig:phase}
\end{figure*}

First we consider the dynamics of a test particle orbiting around the Earth-Moon binary. The left panels of Fig.~\ref{fig:phase} show the $i\cos \phi-i\sin \phi$ phase plane for two different Earth-Moon semi-major axes. 
The $a_{\rm EM}=3\,\rm R_\oplus$ diagram (top left) shows that initial inclinations $i\le 80^\circ$ are prograde circulating orbits.  By running a finer grid we find that with initial  inclinations between $83^\circ$ and $97^\circ$ there are librating orbits.  Librating orbits have their angular momentum vector precessing about the eccentricity vector instead of the angular momentum vector of the Earth-Moon system.  Orbits starting at inclination of $i>99^\circ$  are retrograde circulating orbits.  The angular momentum vector precesses about the negative of the angular momentum vector of the Earth-Moon system.

The top right panel is  the same as the top left but the path of the unit  angular momentum vector of the particle is plotted on the surface of a sphere.  Inside the sphere are shown the unit angular momentum vector for the Earth-Moon system in green as well as the unit eccentricity vector of the Earth-Moon system in red.  The view is chosen to highlight those orbits which librate around the eccentricity vector compared with the path of circulating orbits, shown in blue, which go around the angular momentum vector.

The lower panels are the same as the upper panels but for a larger Earth-Moon semi-major axis of $a_{\rm EM}=15\,\rm R_\oplus$.  This semi-major axis does not support a librating region, and there is a small prograde circulating region and a small retrograde circulating region. Between these two regions, circumbinary particle orbits are unstable to KL oscillations and are ejected.  Some of the orbits in Fig.~\ref{fig:phase} do not repeat exactly, but they are stable for 50,000 Earth-Moon orbital periods. In the next section we discuss the particle stability in more detail. 

\section{Orbital stability of circumbinary Earth-Moon material}
\label{stability}

\begin{figure*}
\includegraphics[width=1.1\columnwidth]{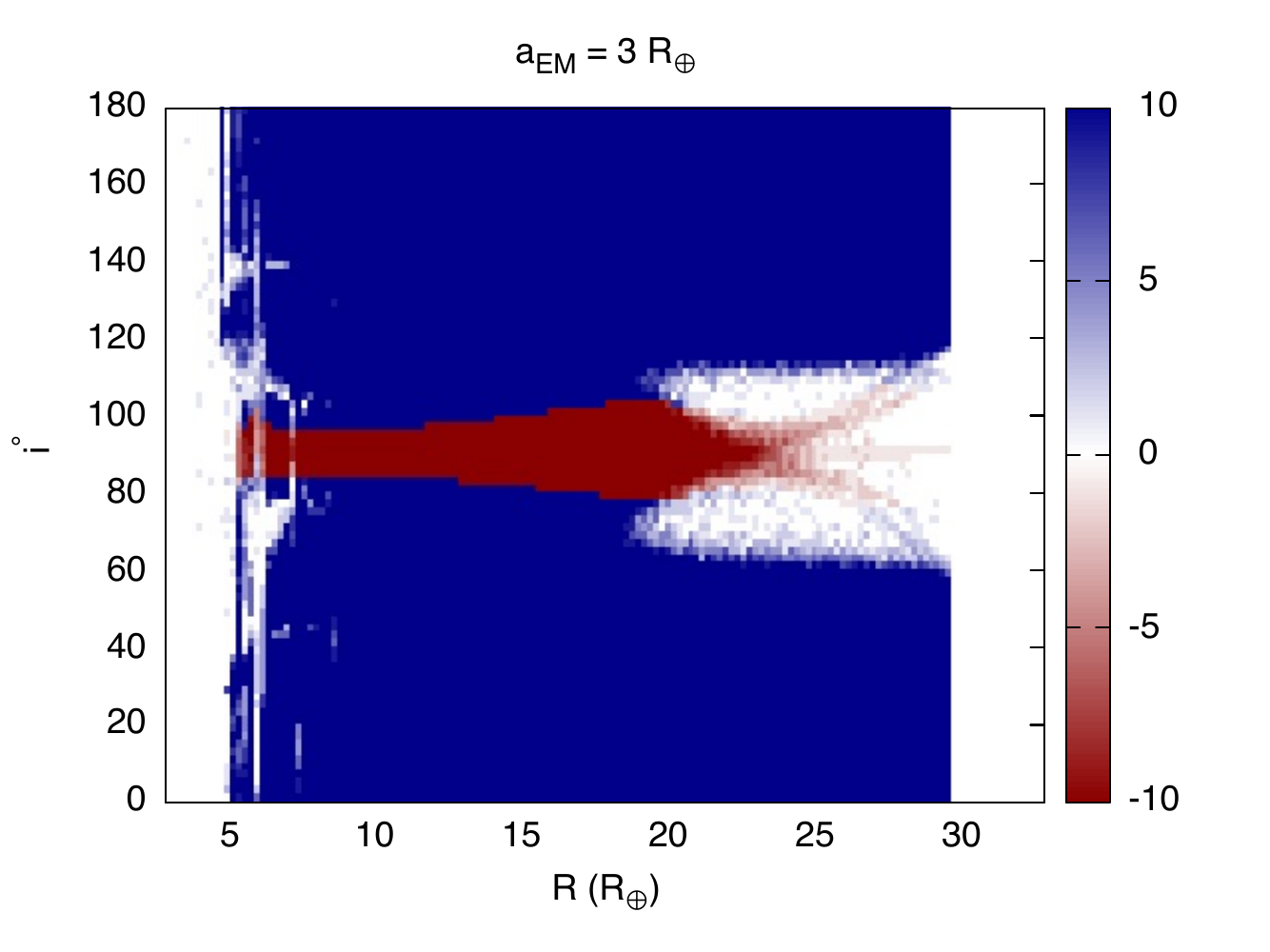}
\hspace{-0.25in}
\includegraphics[width=1.1\columnwidth]{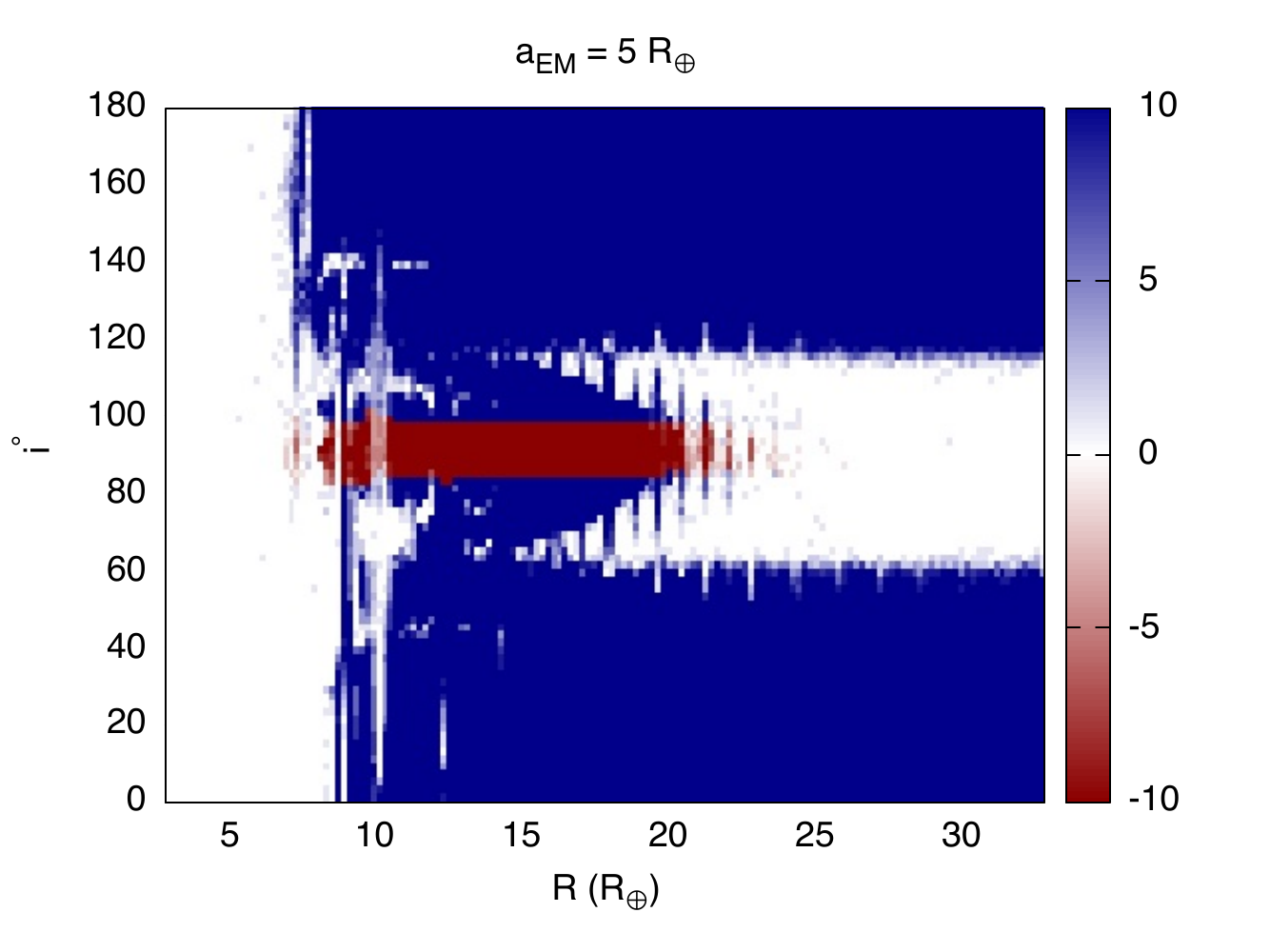}
\includegraphics[width=1.1\columnwidth]{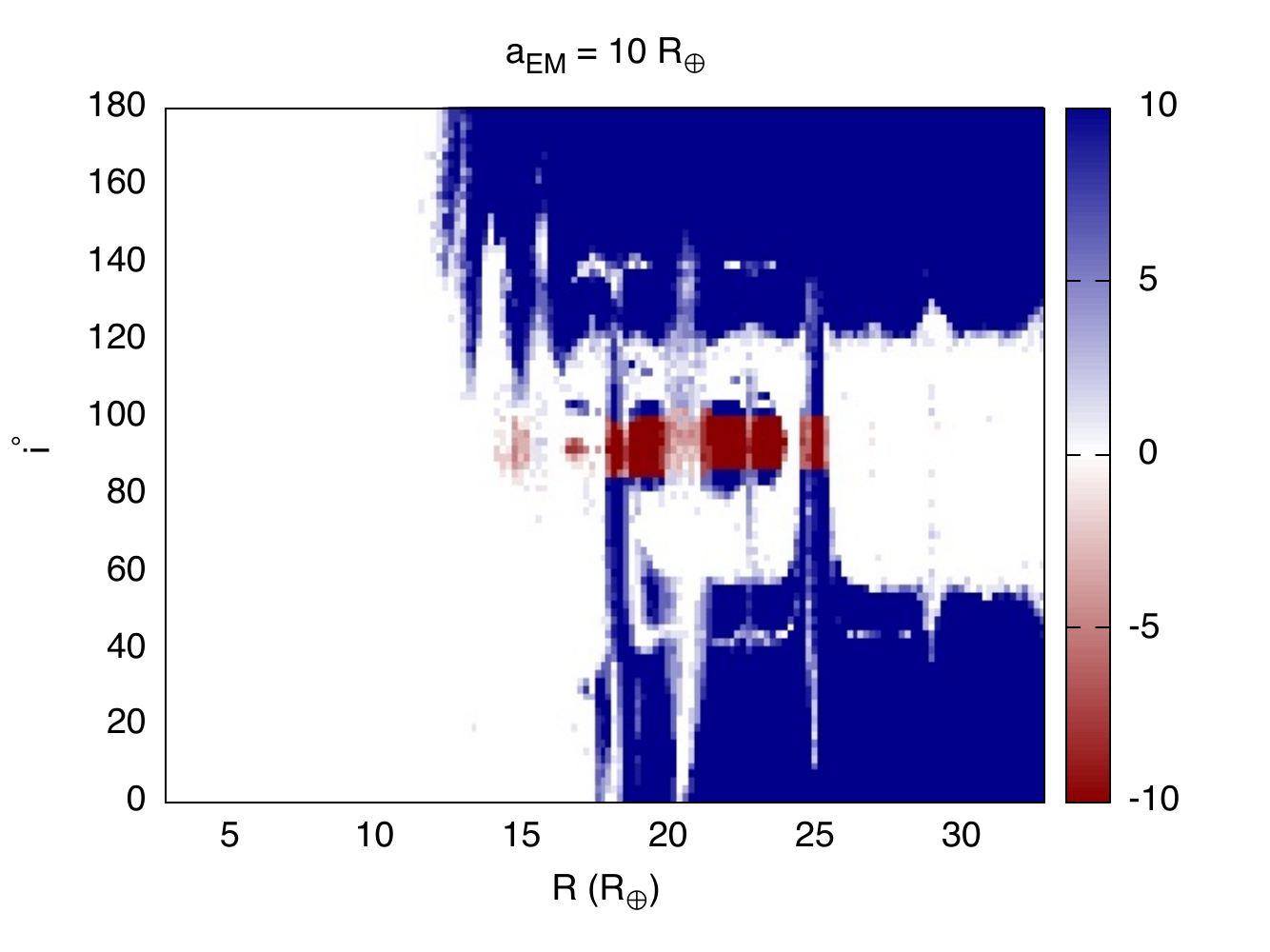}\hspace{-0.25in}
\includegraphics[width=1.1\columnwidth]{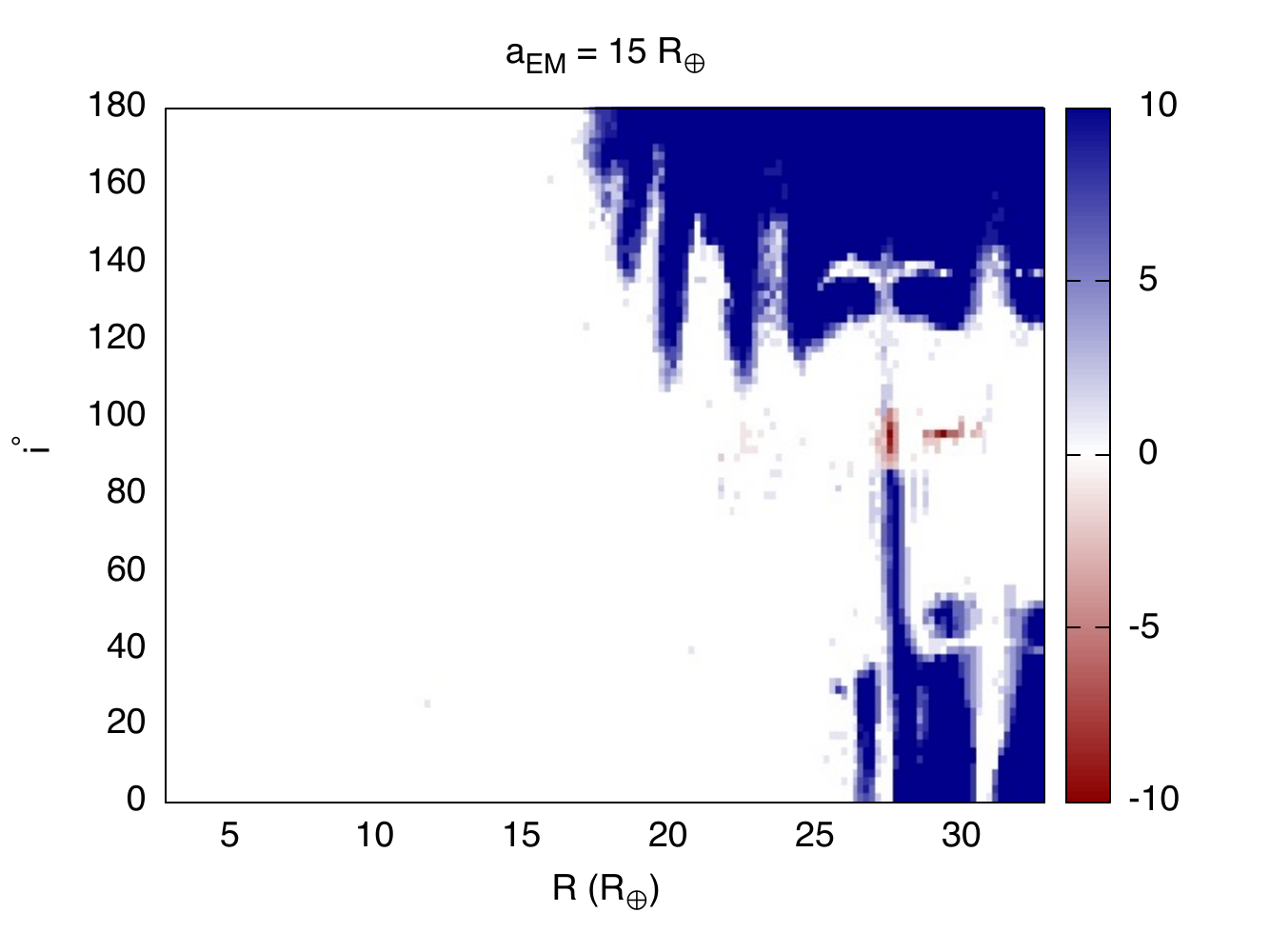}
\caption{Stability diagrams for $a_{\rm EM}=$ 3, 5, 10 and $15\,\rm R_\oplus$. Each pixel shows the results of 10 simulations with different values of the initial true anomaly.  If there is at least one librating orbit, the pixel is colored red.  The intensity of the color shows how many of the 10 orbits were stable.}    
\label{fig:stab}
\end{figure*}

In order to explore the stability of orbits around the Earth-Moon system we set up a grid of simulations with our standard setup.   We consider semi-major axis of the Earth-Moon system  of $a_{\rm EM}=3$, $5$, $10$ and $15\,\rm R_\oplus$ and the current value of $60\,\rm R_\oplus$.

With this setup we make a grid of initial inclinations and initial orbital radius of a test particle.  The inclination steps in 2$^\circ$ increments from 1 up to 179$^\circ$.  The radius  steps in $0.2 \, \rm R_\oplus$ increments starting at the radius of the Earth-Moon system up to a radius beyond which there are no more librating orbits.  Each of these combinations makes a pixel on Fig.~\ref{fig:stab}.  For each combination of $i$ and $R$ we run 10 test particle orbits with different initial true anomalies.  We track each test particle orbit for a time of 50,000 Earth-Moon orbits.  The orbit is labeled as unstable if during this time the eccentricity of the test particle orbit becomes larger than 1, the semimajor axis becomes larger than 10$\,a_{\rm EM}$, or the semimajor axis becomes less than $a_{\rm EM}$ \citep[see for example][]{Quarles2018,Chen2020}.  If the orbit is stable and remains in the upper half of the phase diagram then we label it as a librating orbit.  After the simulation is complete, there are 10 orbits that are either unstable, librating or circulating.  The intensity varies from white (no stable orbits) to either red or blue if all orbits are stable.    The color of the pixel is set to red if at least one of the 10 is a librating orbit and blue if only circulating orbits are found.    The results of this for $a_{\rm EM}=3$, 5, 10 and 15$\,\rm R_\oplus$ are shown in Fig.~\ref{fig:stab}.

At early times when the Earth and Moon were much closer together, there was a small region with stable librating orbits that is seen in the  $a_{\rm EM}=3$ and 5$\,\rm R_\oplus$ plots.   The inner boundary in all cases is determine by test particle orbits being too close to the inner binary \citep[see e.g.][]{Holman1999,Chen2020}.  Following \cite{Martin2022} we can approximate the outer boundary by comparing the timescale for libration to the timescale for KL oscillations.  The KL oscillations \citep{Zeipel1910,Kozai1962,Lidov1962} are driven by  the outer companion and have the test particle cycle through changes in eccentricity and inclination.
The outer boundary occurs when the timescale for KL is equal to the libration timescale for the test particle.  Adopting the equations for the time scale of each from \cite{Martin2022}, we estimate that the stable librating region ends at approximately $R=21 \,\rm R_\oplus$ for  $a_{\rm EM} = 10 \, \rm R_\oplus$ and approximately $R=25 \,\rm R_\oplus$ for  $a_{\rm EM}=15 \, \rm R_\oplus$.  
Since the inner boundary for stable orbits is approximately $R=2\,a_{\rm EM}$, this suggests that a stable librating region will not exist outside of a value for $a_{\rm EM}\approx 10 - 15\, \rm R_\oplus$.  This is in rough agreement with the simulations in Fig.~\ref{fig:stab}, where one can see the librating region is mostly gone by $a_{\rm EM}=15 \, \rm R_\oplus$.

\begin{figure}
\includegraphics[width=\columnwidth]{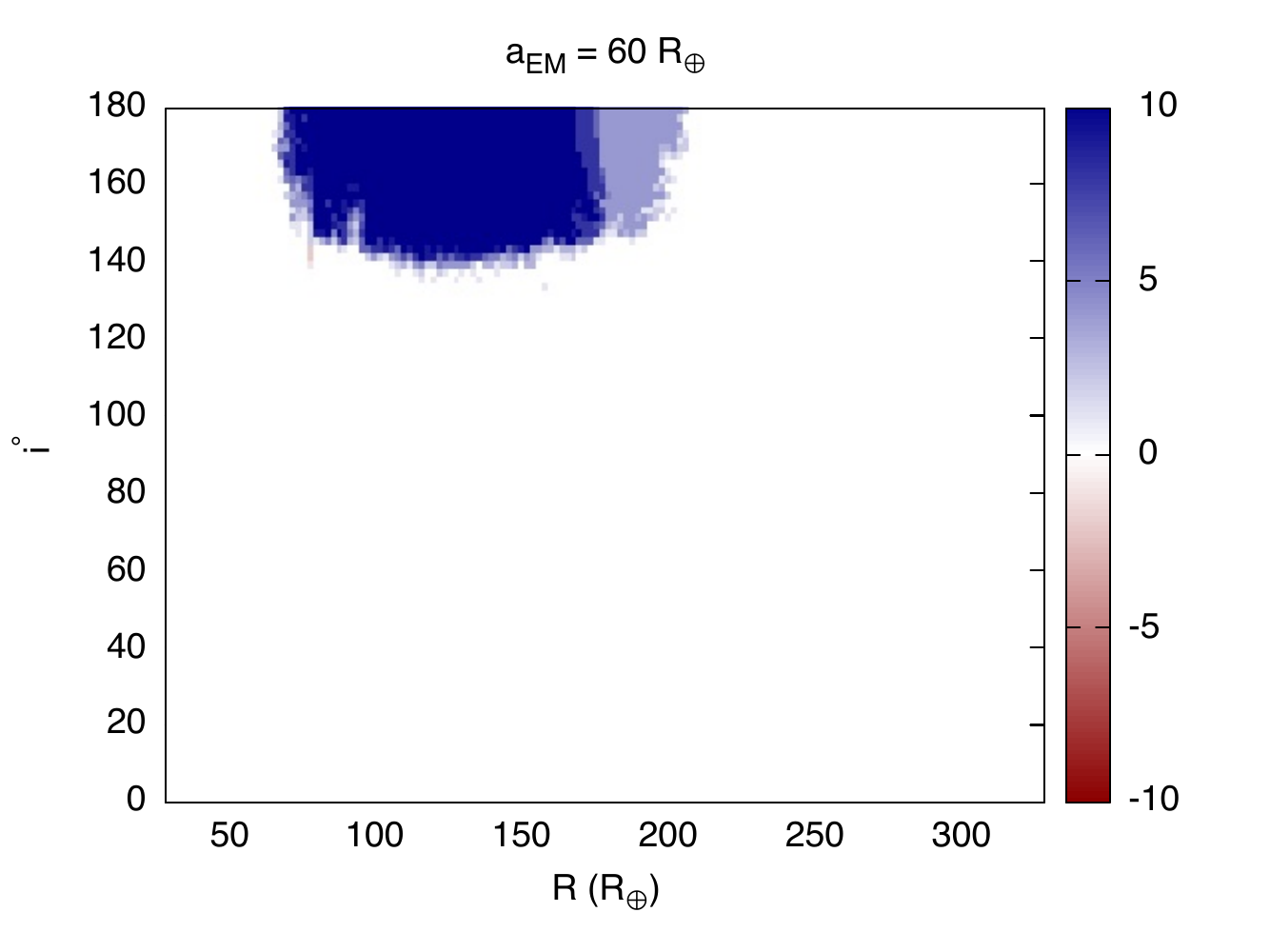}
    \caption{Same as Fig.~\ref{fig:stab} but with the current Earth-Moon semi-major axis of $a_{\rm EM}=60\,\rm R_\oplus$.}
    \label{fig:stabnow}
    \end{figure}

 For semi-major axis $a_{\rm EM}\gtrsim 15 \, \rm R_\oplus$, there is not a librating region around the polar orbits.   This is shown in Fig.~\ref{fig:stabnow} which has  $a_{\rm EM}= 60 \, \rm R_\oplus$. For the current Earth-Moon system, nearly all circumbinary orbits are unstable but there is a small retrograde stable region.  There is no stable librating region in the current earth moon configuration.

\begin{figure*}
\includegraphics[width=1.1\columnwidth]{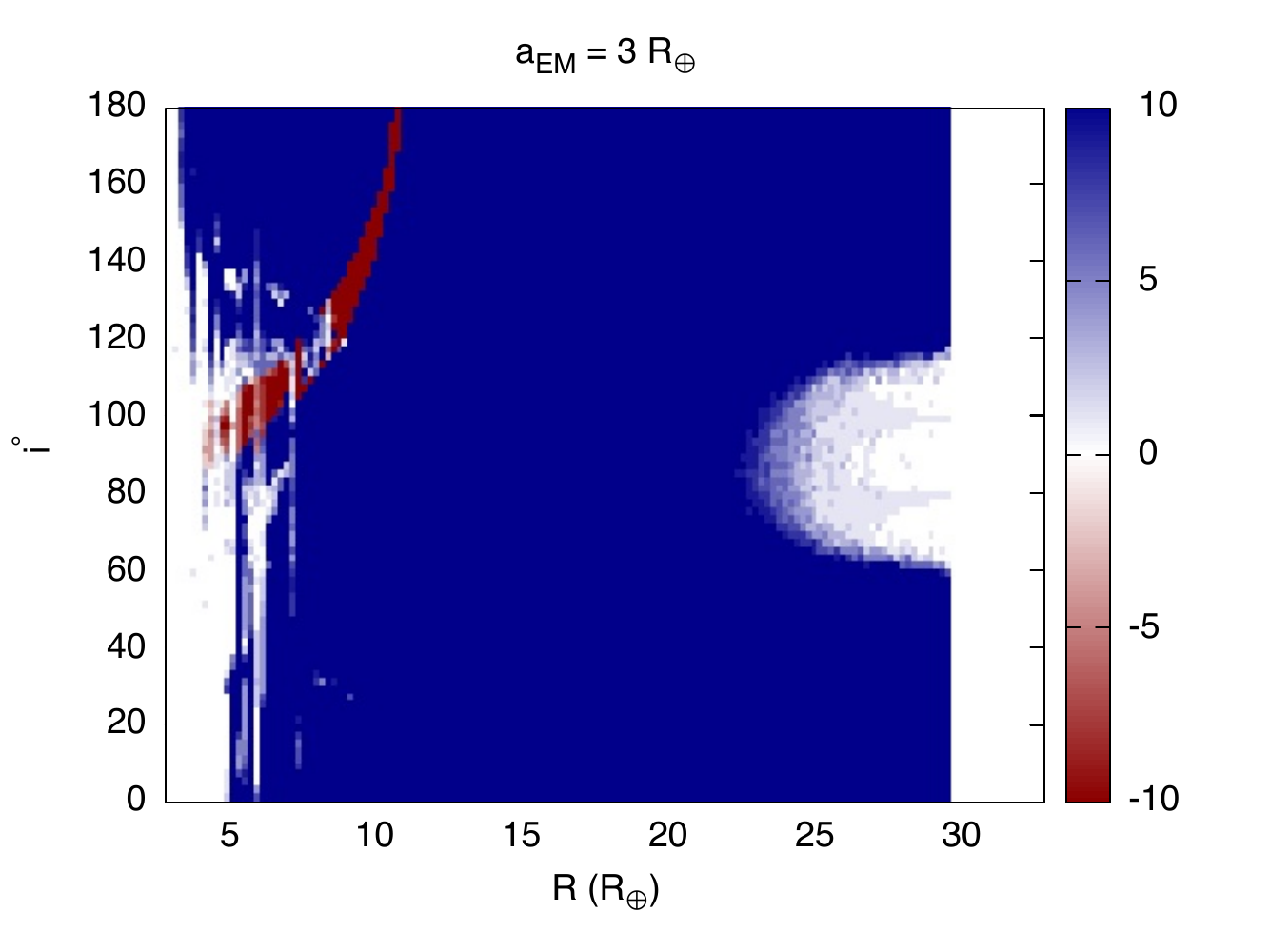}
\hspace{-0.25in}
\includegraphics[width=1.1\columnwidth]{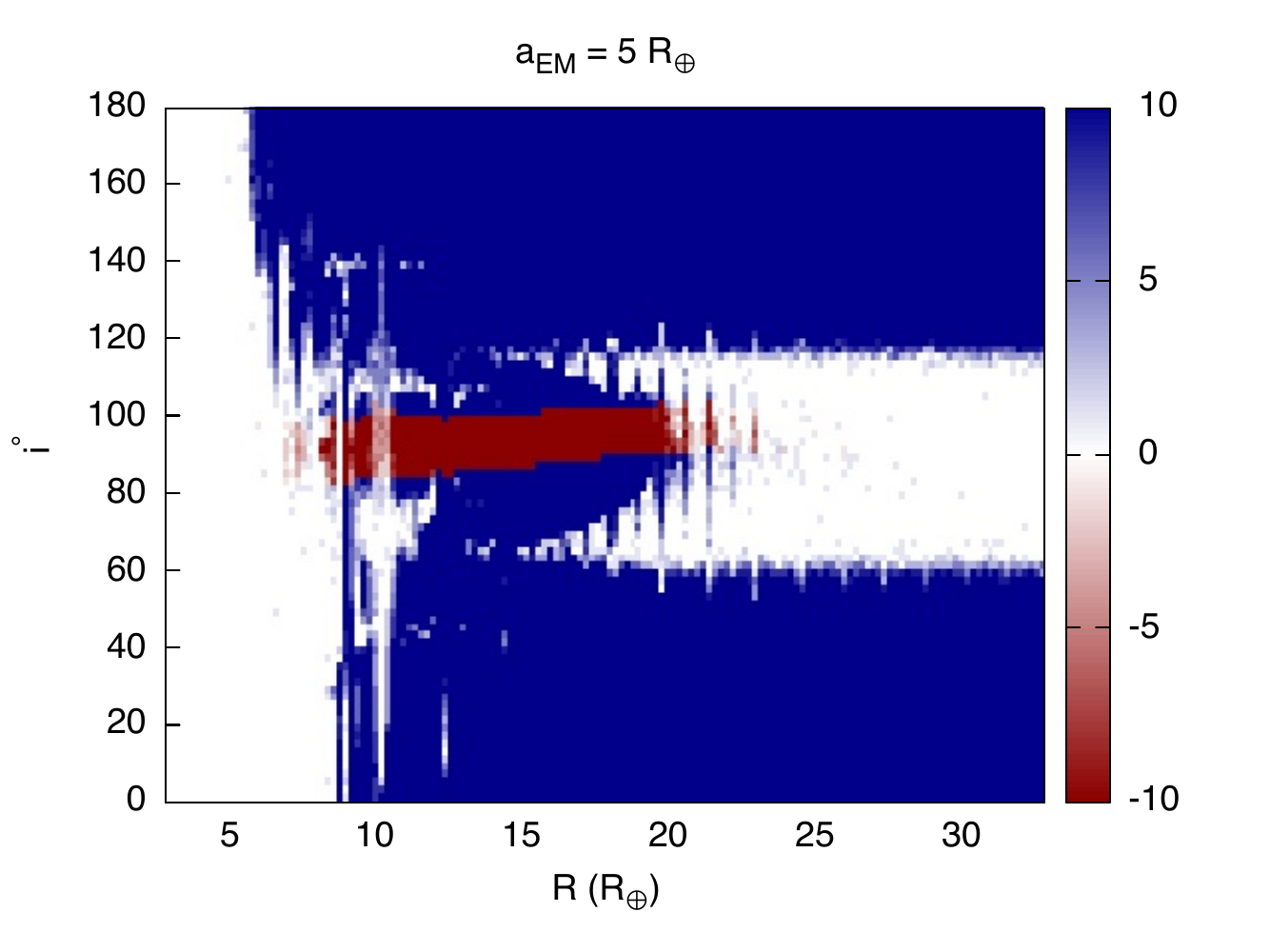}
\includegraphics[width=1.1\columnwidth]{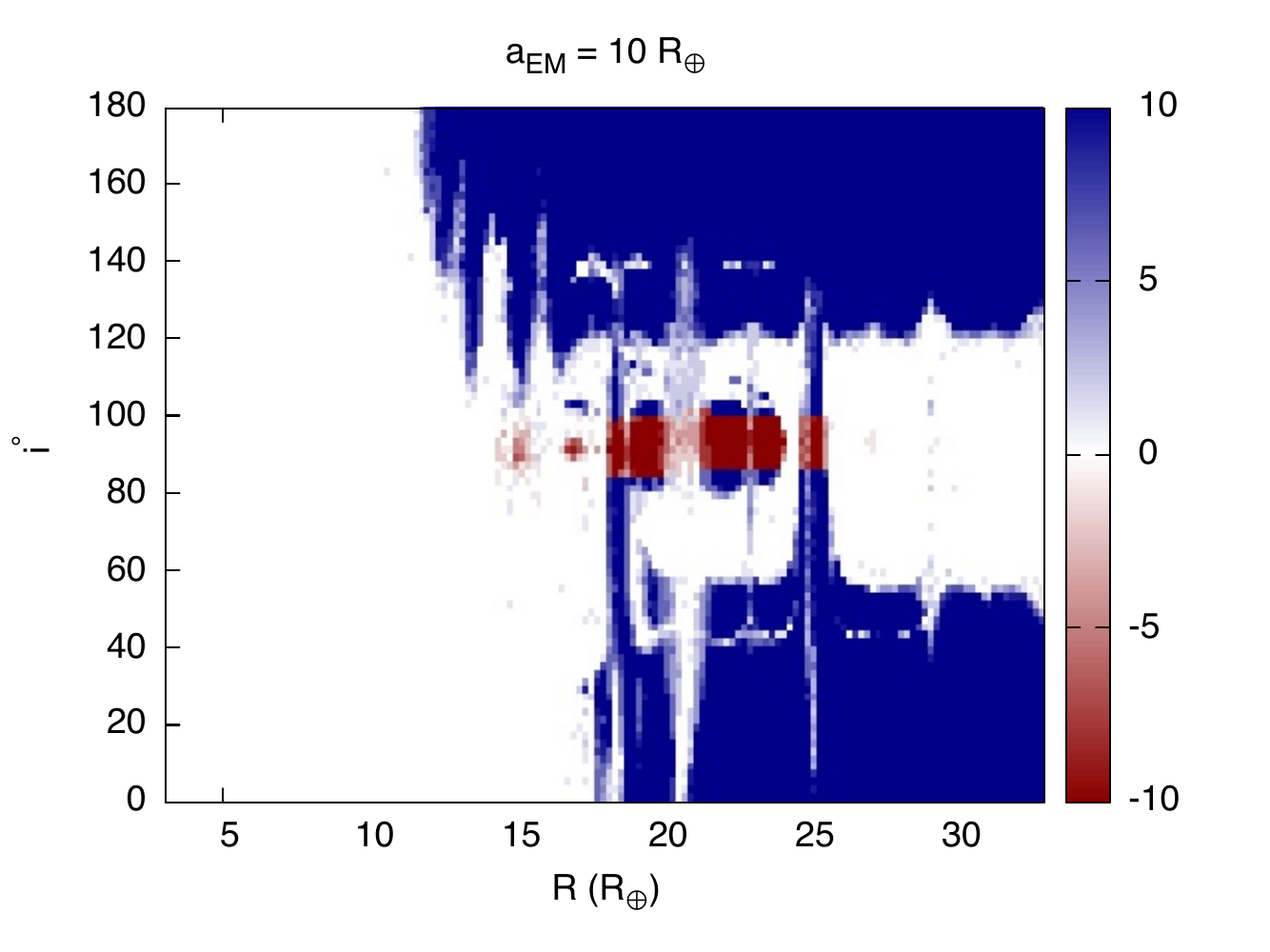}
\hspace{-0.25in}
\includegraphics[width=1.1\columnwidth]{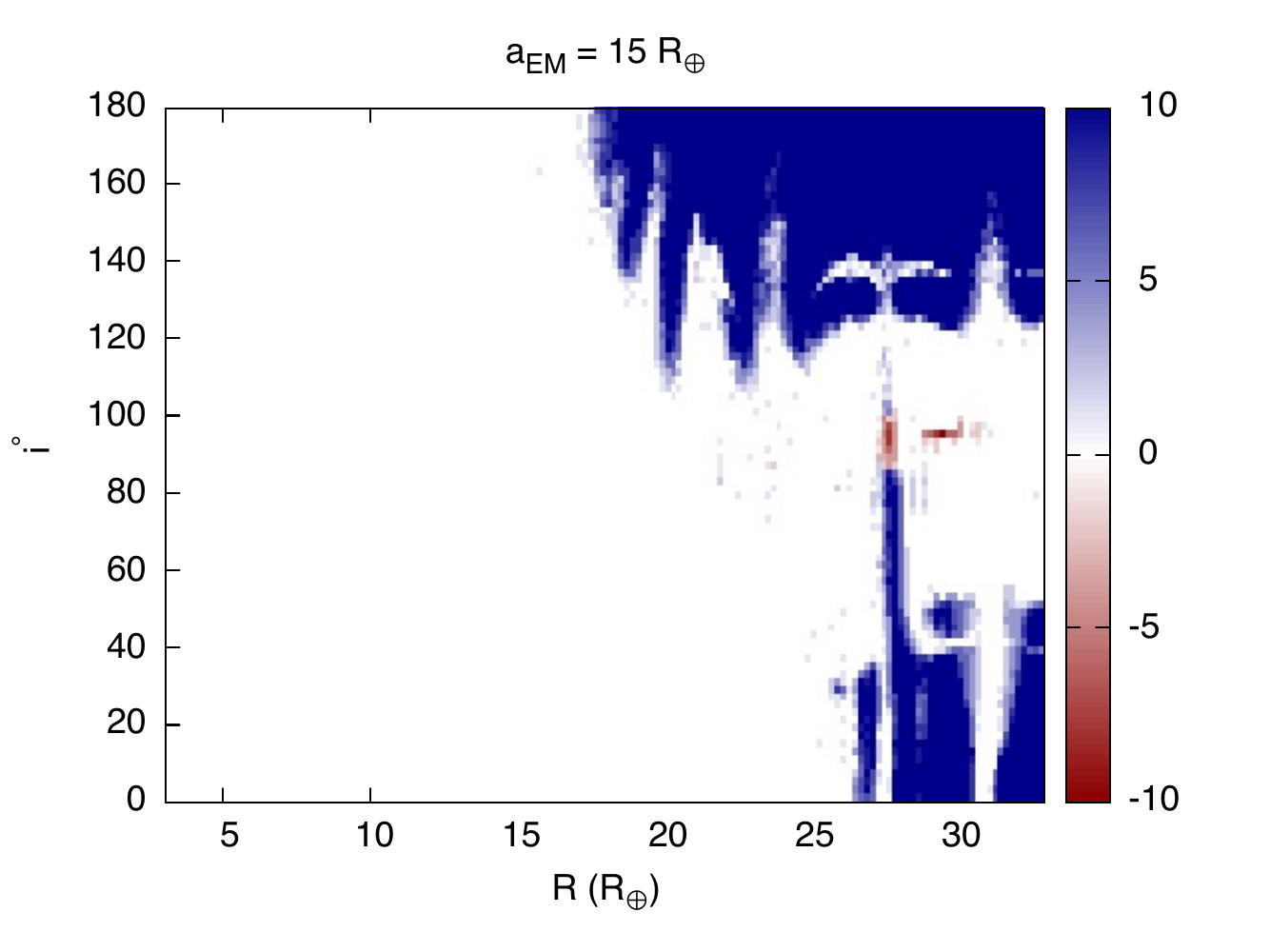}
    \caption{Same as Fig.~\ref{fig:stab} except tides are included.}
    \label{fig:stabtide}
\end{figure*}

\section{Tides and Tidal Dissipation}
\label{tides}

In order to determine the effects of tides on the librating orbit region we reran the simulations with tidal forces included using the {\sc rebound}x module for constant time lag tides \citep{Baronett2022}.  This adds a constant time lag tides as described in \cite{Hut1981}.  Tidal forces are likely to have been weaker at early times after the formation of the moon compared to the current day.  We include calculations with the current tidal parameters to place an upper limit on the effect of tidal forces.   We obtained the parameters from one of the implementation examples for \cite{Lu2023} in the {\sc rebound}x library. 
For non-dissipative tides, we take the radius of Earth to be $4.264\times 10^{-5}\,\rm au$ and the potential Love number of degree 2 to be $k_2= 0.298$.  
  This component of the tidal potential conserves energy but causes the binary to precess.   We  also consider dissipative tides by adding   constant time lag, $\Omega=2\pi /\rm day$.  We found  that the dissipation was unimportant in the librating region as the libration time scales are much shorter than the tidal dissipation time scale.  The precession of the orbit did have an important effect however.

Similar to apsidal precession caused by general relativity \citep{Zanardi2018, Lepp2022,Zanardi2023} and that caused by a companion \citep{Lepp2023}, the prograde precession of the eccentricity vector through tides causes the stationary inclination of the circumbinary particles to increase from 90$^\circ$.  This in turn causes the librating region to move to higher inclinations.  

The simulations in Fig~\ref{fig:stab} were all rerun with non-dissipative tides and the results are shown in Fig.~\ref{fig:stabtide}.  The stability and orbit characterization for the $a_{\rm EM}=10$ and $15 \, \rm R_\oplus$ are mostly unchanged and look very much the same as without tides.  However in the $a_{\rm EM}=5 \, \rm R_\oplus$ panel one can clearly see the libration region shifting to higher inclinations with for larger orbital radii.  For the $a_{\rm EM}=3 \, \rm R_\oplus$ case the effect is even more dramatic.  Here we see the librating region very rapidly shifting to higher inclination and the librating region reaches $i=180^\circ$ by $R\approx 10 \, \rm R_\oplus$.  This is a dramatic example of precession of a binary changing where the stationary inclination is and removing the possibility for librating orbits.
Even with tides, there is a significant libration region for  $a_{\rm EM}=5\,\rm R_\oplus$ and a smaller region at $a_{\rm EM}=10\,\rm R_\oplus$  that still poses stable librating regions.

\section{Discussion and Conclusions}
\label{conc}

We have shown that stable polar circumbinary orbits existed around the Earth-Moon binary system after the formation of the Moon. As the separation of the Earth-Moon increased through tides, the region of space for which the polar orbits exists decreased. The current day Earth-Moon system does not have any possible stable polar orbits since the nodal precession driven by the Sun dominates that driven by the Earth-Moon binary at all orbital radii. 

Polar circumbinary material can drive eccentricity growth of the binary \citep{Farago2010,Chen2019,Martin2019,Martin2024}. If a significant mass of material ended up on polar or librating orbits, then the eccentricity of the Earth-Moon binary could have been increased as a result of its interaction.  

The existence of stable polar orbits around the young Earth-Moon system has implications for exoplanets. The range of initial inclinations for which librating orbits exist increases with the binary eccentricity \citep{Doolin2011}. Therefore, a more highly eccentric planet-moon orbit is more likely to able to host polar circumbinary material. Significant polar orbiting material could form a second moon  in a polar orbit around the planet-moon inner binary.
A  polar orbiting moon would drive retrograde apsidal precession of the planet-moon inner binary \citep[see also, e.g.][]{Zhang2019,Baycroft2023}.

\vskip .2in
\section*{Acknowledgements}
We thank Scott Tremaine for suggesting the problem.   We acknowledge support from NASA through grant 80NSSC21K0395.
    

\bibliographystyle{aasjournal}
\bibliography{ct} 

\end{document}